\documentstyle[12pt]{article}

\newcommand{\bpi}{\mbox{\boldmath $\pi$}}
\makeatletter
\def\agt{\compoundrel >\over\sim}
\def\alt{\compoundrel <\over\sim}
\def\compoundrel#1\over#2{\mathpalette\compoundreL{{#1}\over{#2}}}
\def\compoundreL#1#2{\compoundREL#1#2}
\def\compoundREL#1#2\over#3{\mathrel
    {\vcenter{\hbox{$\m@th\buildrel{#1#2}\over{#1#3}$}}}}
\makeatother
\begin{document}
\begin{center}
{\bf Weak chiral lagrangian and\\
$\pi^+$ emission in ${}^{4}_{\Lambda}$He}\\[0.15cm]
M.Shmatikov\\
{\it Russian Research Center ``Kurchatov Institute'',
123182 Moscow, Russia}\\[0.5cm]
{\bf Abstract}\\[0.15cm]
\end{center}
A new mechanism of  $\pi^+$-meson emission from the
${}^{4}_{\Lambda}$He hypernucleus is suggested. It consists of a
weak $\Lambda\rightarrow n + \pi^{+}+\pi^{-}$ decay followed by
$\pi^{-}$ absorption by the hypernuclear core. The contribution of
this mechanism is compared to that of the ''standard'' one 
($\Lambda\rightarrow n +\pi^{0}$ decay followed by the
$\pi^{0}\rightarrow\pi^{+}$ charge exchange on the core).
Uncertainties of the estimated effect are discussed.\\[0.15cm]
PACS numbers: 12.39.Fe,14.20.Jn,21.80.+a
\newpage
\section{Introduction}
The puzzle of $\pi^{+}$ emission in the (weak) ${}^{4}_{\Lambda}$He hypernucleus decay remains 
unsolved yet. Experiments \cite{Bohm,Keyes} have shown that the ratio $R$ of $\pi^+$ and
$\pi^-$ yields which is defined as
\begin{equation}
R(\pi^+/\pi^-)=\frac{\Gamma({}^4_{\Lambda}{\rm He}\rightarrow \pi^+ + X)}
{\Gamma({}^4_{\Lambda}{\rm He}\rightarrow \pi^- + X)}
\end{equation}
equals
\begin{equation}
R(\pi^+/\pi^-)\approx (5\pm 2)\%
\label{exp}
\end{equation}

Since the free $\Lambda$-hyperon decays with the emission of $\pi^+$ and $\pi^0$ mesons only,
the $\pi^+$ mode necessarily involves core's nucleon(s). Investigation of the $\pi^+$ emission
mechanisms was pioneered in \cite{Dalitz}, where a two step mechanism was considered. 
It involves
the $\Lambda\rightarrow n +\pi^0$ decay followed by the charge exchange
\begin{equation}
\Lambda p\rightarrow\pi^0 n p\rightarrow \pi^+ nn
\end{equation}
The $R$ ratio with such mechanism proved to be much below the experimental
value (\ref{exp}). Other mechanisms (mainly strong $\Lambda p\rightarrow\Sigma^+ n$ conversion
followed by the $\Sigma^+\rightarrow \pi^+ + n$ decay) were shown \cite{Hippel} to give
negligible ($\approx 0.2\div 0.4$\%) contribution.
The two-step charge-exchange approach was developed in \cite{Gal} using more precise and 
up-to-date input parameters. The calculated branching ratio is about two times larger than in
\cite{Dalitz} $R\approx (1.22\div 1.29)\%$ but still is about two standard deviations below
the experimental value.

Conclusions of \cite{Hippel} concerning negligible contribution of the $\Sigma^+$
involving decay mechanism were re-examined in \cite{Gibson}. It was argued that the
correct account of the $\Lambda-\Sigma$ conversion, which is essential for hypernuclear
binding, may yield the $R$ ratio of the order of 5\%. The $R$ value in this approach is
scaled by the admixture $P_{\Sigma}$ of the $\Sigma$-hyperon in the hypernuclear wave function. 
The experimental value of $R$ can be reproduced with $P_{\Sigma}\approx 9\%$.

Alternative approach to the problem was developed in \cite{Oka}. The $\pi^+$ emission
in a hypernuclear decay was related to the $\Delta I = 3/2$ component of the weak nonleptonic
lagrangian. Corresponding amplitude was calculated in the quark model. The suggested approach 
does not take into consideration the yield of $\pi^-$ mesons from the decay of (almost)
free $\Lambda$ in the hypernucleus so that the obtained results cannot be directly confronted
to the experimental value (\ref{exp}).

Mechanisms invoked for explaining the $\pi^+$ decay branching ratio have one feature in common.
The $\pi$-meson is emitted either in a weak vertex ($\Lambda\rightarrow n+\pi^0$ \cite{Gal}
or $\Sigma^+\rightarrow n+\pi^+$ \cite{Gibson}) or in a strong one ($\pi^+$ emission
by protons prior and after the $\Lambda N\rightarrow NN$ process \cite{Oka}). We suggest a
mechanism involving weak {\it scattering} of the $\pi$-meson or, stated differently, the
mechanism of emission of {\it two} $\pi$-mesons. To this end we consider the structure
of the weak interaction hadronic lagrangian.
\section{Weak interaction lagrangian}
The (dominant) octet component of the lowest-order $\Delta S = 1$ chiral lagrangian is given 
by \cite{Donoghue}
\begin{equation}
{\cal L}^{\Delta S = 1} = h_D{\rm Tr}\overline{B}\left\{\xi^\dagger h\xi,B\right\}+
h_F{\rm Tr}\overline{B}\left [\xi^\dagger h\xi,B\right ]\; ,
\label{lagran}
\end{equation}
where $h_D$ and $h_F$ are numerical constants and the $h$ matrix 
\[
h= \left(
\begin{array}{ccc}
0 & 0 &0\\
0 & 0 &1\\
0 & 0 &0 
\end{array} 
\right)
\]
selects the $\Delta S = 1$ transitions. $B$ denotes the octet-baryon matrix and the 
$\xi$ matrix reads
\begin{equation}
\xi = {\rm e}^{i\,\pi/f}
\end{equation}
where $f$ is the $\pi$-meson decay coupling constant and $\pi$ is the matrix of
the pseudoscalar meson octet
\begin{equation}
\xi = \frac{1}{\sqrt{2}}\,\left(
\begin{array}{ccc}
\pi^0/\sqrt{2} + \eta/\sqrt{6}& \pi^+& K^+\\
\pi^-& -\pi^0/\sqrt{2} +\eta/\sqrt{6}& K^0\\
K^-& \bar{K}^0& -2\,\eta/\sqrt{6}
\end{array}
\right)
\end{equation}
The lagrangian (\ref{lagran}) contains vertices of the $\Delta S = 1$ decays with
any number of the emitted pions. Expansion of the lagrangian up to linear in the $\pi$-meson
field terms yields amplitudes of the nonleptonic hyperon decays. Comparison of the latter
to the experimental values of corresponding amplitudes allows determining the $h_{D,F}$
coupling constants so that the lagrangian actually does not contain any unknown parameters.

It is instructive to consider qualitative features of the lagrangian (\ref{lagran}) and, 
especially, that of the $\Lambda \rightarrow N +\pi+\pi$ vertex. (Such a decay mode of the free
$\Lambda$-hyperon is forbidden kinematically, however, it may occur in a nuclear
environment). 
\begin{itemize}
\begin{enumerate}
\item Absence of derivative couplings implies that higher order terms in mesonic 
fields
do not involve any suppression factor related to small $\pi$-meson momenta as compared
to the $\Lambda \rightarrow N +\pi$ amplitude. 
\item  The isotopic 
structure of the lagrangian shows that a pair of $\pi$-mesons is emitted in the $I=0$ state. 
\item The 
$\pi$-mesons are produced in the S-wave state, and parity is conserved in the vertex.
\end{enumerate}
\end{itemize}
Thus the weak chiral lagrangian yields the $\Lambda \rightarrow n + \pi +\pi$ vertex
reading (wave functions of baryons are omitted for notational brevity)
\begin{equation}
M^{\pi\pi} =-\frac{{\bpi}{\bpi}}{2\,f^2}
\label{twopi}
\end{equation}
For comparison, the amplitude of the ``benchmark'' $\Lambda \rightarrow n + \pi $ decay
in the same normalization has the form
\begin{equation}
M^{\pi} = i\,\frac{\pi^0}{f}
\label{onepi}
\end{equation}
Absorption of the $\pi^-$-meson emitted in the $\Lambda\rightarrow n +\pi^+ + \pi^-$ vertex (\ref{twopi})
by the core of a hypernucleus gives rise to the decay mechanism depicted in Fig.1. Hereafter
we shall call it an ``absorption'' mechanism.
\section{Estimate of the absorption mechanism contribution }
The mechanism of $\pi^+$ production considered in \cite{Dalitz,Gal} (the
$\pi^0 + p\rightarrow \pi^+ + n$ charge-exchange) is a microscopic one-nucleon description 
of the more generic process shown in Fig.2.
To make estimates of the absorption mechanism contribution we compare it to that of the
latter. Indeed, the vertex of the $\Lambda\rightarrow n +\pi^+ + \pi^-$ decay may
be considered as an amplitude of the weak strangeness changing 
$\Lambda + \pi^+\rightarrow n +\pi^+$ scattering. The charge exchange (CEX) and the absorption
(ABS) mechanisms have similar structure: emission of the $\pi$-meson (in the weak and the strong
interaction vertex respectively), its propagation and scattering (on the core and on the
$\Lambda$-hyperon respectively). The ratio of corresponding amplitudes
\begin{equation}
\rho \equiv \frac{A^{\rm abs}}{A^{\rm cex}}
\end{equation}
has the form
\begin{equation}
\rho = \frac{M^{\pi\pi}\,D_{\pi}\,M^{\rm abs}}{M^{\pi}\,D_{\pi}\,M^{\rm cex}}\;,
\label{ratio}
\end{equation}
where $D_{\pi}$ is the $\pi$-meson propagator, $M^{\pi\pi}$ and $M^{\pi}$ are the vertices
of the $\Lambda\rightarrow n +\pi+\pi$ and $\Lambda\rightarrow n +\pi$ decays respectively. The
$M^{\rm abs}$ is the amplitude of the $\pi^- + A\rightarrow X$ absorption by the hypernuclear
core and $M^{\rm cex}$ is that of the $\pi^0 + A\rightarrow \pi^+ + X$ charge exchange on the
core (in the   ${}^{4}_{\Lambda}$He case the core is the ${}^3$He nucleus).

The absorption and the charge-exchange processes differ mainly by kinematics. In the latter
case the emitted $\pi$-meson is on-shell with the kinetic energy ${\rm T}_{\pi}\approx 30$~MeV. It
implies that the $\pi$-meson propagation between the emission and the charge-exchange
on the core is described by the imaginary part of the $D_{\pi}$ propagator or, in the coordinate
space, by the outgoing spherical wave $\sim\, {\rm e}^{iqr}/r$ (where $q$ is the $\pi$-meson
momentum). At the same time the $\pi^-$-meson emitted in the $\Lambda\rightarrow n +\pi^++\pi^-$
vertex is off-shell. Indeed, its mass squared $q^2$ is related to the effective mass
$\mu$ of the neutron and $\pi^+$ pair:
\begin{equation}
q^2 = (m_{\Lambda} - \mu)^2
\label{mass}
\end{equation}
The $q^2$ value controlling the $\pi$-meson virtuality attains its maximum at the minimal
$\mu_{\rm min}$ which is equal to $\mu_{\rm min}= m_N + m_\pi$ yielding:
\begin{equation}
q^2 - m^2_{\pi} \alt \left[m_\Lambda - (m_N + 2m_\pi)\right]\,(m_\Lambda - m_N)
\end{equation}
or
\begin{equation}
q^2 - m^2_{\pi} \alt - m^2_{\pi}
\label{off}
\end{equation}
The latter equation shows that the $\pi^-$-meson in the absorption mechanism can propagate
effectively at the distances  $\approx 1/m_{\pi}$ in contrast to the charge exchange mechanism where
the $\pi$-meson, being on-shell, propagates at arbitrary large distances from the source.

This circumstance may prove to be crucial for the magnitude of the absorption-mechanism effect.
Indeed, the mechanisms depicted graphically in Figs.1 and 2 yield the weak-interaction
amplitude ${\cal A}^W$. Baryons in the initial and final state are subject to strong
interaction so that the total amplitude of the process reads
\begin{equation}
{\cal M} = \int\, \psi^{*}_{f}\,{\cal A}^W\psi_i\,d\tau
\end{equation}
where $\psi_{i,f}$ are the wave functions controlled by the initial- and final state 
interaction and $d\tau$ is the phase-space volume. As it was stated above the absorption 
mechanism contribution is saturated at distances $\sim 1/m_\pi$ (though formally all distances 
contribute to the charge-exchange mechanism it is also controlled by the small separation
region - see \cite{Gal}). The behavior of the $\psi_i$ wave function is governed by the
initial state interaction between the $\Lambda$-hyperon and the nuclear core.The situation 
with the latter is rather controversial. In \cite{Dalitz} a $\Lambda$-core wave function
was used corresponding to a repulsion at small separations. It was criticized in \cite{Gal} 
where more realistic wave function was proposed. The latter has no pronounced repulsion
at small distances and vanishes at distances $\approx 2$~fm. With this $\Lambda$-core
wave function the absorption mechanism does not involve additional suppression as compared
to the charge-exchange process. It implies that the difference in kinematics between two
processes which reflects in different $q^2$ values in the $\pi$-meson propagators in 
(\ref{ratio}) does not affect the values of corresponding amplitudes. As a result the
ratio of the $\pi$-meson propagators drops out from the (\ref{ratio}) ratio. Taking into
account that the ratio of the weak-interaction amplitudes $M^{\pi\pi}/M^{\pi}\approx 1$ 
(cf. (\ref{twopi}) and (\ref{onepi})) we can rewrite the ratio of two amplitudes in a
simple form
\begin{equation}
\rho \approx \frac{M^{\rm abs}}{M^{\rm cex}}\;,
\label{ratio1}
\end{equation}
The relative contribution of two mechanisms is controlled by the ratio
of the $\pi$-meson absorption and charge-exchange on the hypernuclear core. The expression
of $\rho$ with the amplitudes' relative phase being omitted can be rewritten in a more
transparent form
\begin{equation}
\rho \approx \sqrt{\frac{\sigma^{\rm abs}}{\sigma^{\rm cex}}}
\label{cross}
\end{equation}
\section{Pion-core interaction}
The denominator of the r.h.s. of (\ref{cross}) contains the cross section of the
charge-exchange process
\begin{equation}
\pi^0 + A\rightarrow \pi^+ + X
\end{equation} or, in the considered case of the ${}^{4}_\Lambda$He nucleus
\begin{equation}
\pi^0 + {}^3{\rm He}\rightarrow\pi^+ +\left\{
\begin{array}{l}
p + n + n\\
d + n\\
{}^3{\rm H} 
\end{array}\right .
\label{pia}
\end{equation}
which cannot be measured experimentally. Nevertheless some estimates of the required cross
section can be made on the basis of available experimental data. 

The energy of the $\pi$-meson emitted in the $\Lambda$-hyperon decay is rather small
(${\rm T}_{\rm lab}\approx 30$~MeV). Because of smallness of the phase-space volume it is natural to 
assume then that the charge-exchange
process (\ref{pia}) is dominated by the channel with the minimum number of particles
in final state, i.e. $\pi^0 + {}^3{\rm He}\rightarrow\pi^+ + {}^3{\rm H}$. The cross section
$\sigma^{\rm cex}$ of the latter process can be related, using the isotopic relations, 
to the cross sections of the
elastic $\pi$-meson scattering. Indeed, denoting the cross sections of the
$\pi^{\pm} + {}^3{\rm He}\rightarrow\pi^{\pm} + {}^3{\rm He}$ scattering as $\sigma^{\pm}$ 
one can show that the following relation holds:
\begin{equation}
\sigma^{\rm cex} = \sigma^{+}/6 + \sigma^{-}/2
\label{isa}
\end{equation}

Experimentally at ${\rm T}_{\rm lab}= 98.5$~MeV the cross sections are equal to 
$\sigma^{+}= 66.6\pm 6.7$~mb \cite{sp} and $\sigma^{-}= 40.0\pm 6.0$~mb \cite{sp}.
Substituting these values to (\ref{isa}) we get  $\sigma^{\rm cex}\approx 30$~mb.
To obtain the value of the cross section at the required energy ${\rm T}_{\rm lab}\approx 30$~MeV
we note that it is far below the resonance region. It is natural to assume then that
the matrix element of the process exhibits weak energy dependence and the cross section is controlled
mainly by the available phase space volume. For a nonrelativistic two-particle final state
the latter is scaled by the c.m.s. momentum of the outgoing particles. It implies that the values
of the cross section at ${\rm T}_{\rm lab}\approx 98$~MeV and $\approx 30$~MeV are related by
the $\sqrt{98.5/30}\approx 1.8$ factor. The estimate for the 
$\pi^0 + {}^3{\rm He}\rightarrow\pi^+ X$
charge exchange cross section $\sigma^{\rm cex}$ at the energy corresponding to
the free $\Lambda$-hyperon decay reads:
\begin{equation}
\sigma^{\rm cex}\approx 17\;{\rm mb}
\label{ss}
\end{equation}

The value of $\sigma^{\rm abs}$ in the r.h.s. of (\ref{cross}) cannot be determined directly
from the experiment since the absorbed $\pi$-meson is off-shell. Assuming that this
circumstance does not affect strongly the cross section value we adopt its low-energy
value. At ${\rm T}_{\rm lab}= 64$~MeV the cross section of $\pi^{-}$ absorption by ${}^3$He
equals $\sigma^{\rm abs}=14.7\pm 2.6$~mb \cite{abs}. In absence of kinematical restraints
and beyond the resonance region it is natural to assume weak energy dependence of the
absorption cross section. Substituting the $\sigma^{\rm abs}$ and $\sigma^{\rm cex}$ (\ref{ss}) 
values  in (\ref{cross}) we arrive at the estimate:
\begin{equation}
\frac{M^{\pi\pi}}{M^{\pi}}\alt 1\,.
\end{equation}
It implies that the mechanism generated by the $\Lambda\rightarrow N +\pi+\pi$ vertex
(Fig.1) gives the contribution to the yield of $\pi^+$-mesons in the ${}^{4}_{\Lambda}$He
decay comparable to that of the ``benchmark'' process (Fig.2).

Let us consider now the issue of possible interference of two contributions. Generically the particle composition
of final states produced by the ``absorption'' and ``charge-exchange'' mechanisms are the same so that they
might interfere. However, the kinematics of the final state is rater different. Absorption of the $\pi$-meson
involves at least two nucleons. It is, strictly speaking, true for the on-shell meson, however, one can expect
that the same situation will survive for a slightly off-shell $\pi^-$-meson in the considered case. At the same
time one can argue that  ``charge-exchange'' of  the low-energy pion proceeds predominantly on only one nucleon.
It implies that the interference between two mechanism will be very weak and their contributions will add
incoherently.
\section{Other light hypernuclei}
The mechanism of the $\pi^+$-emission in ${}^{4}_{\Lambda}$He generated by the $\Lambda\rightarrow n+\pi+\pi$
vertex was shown above to give the contribution comparable to that of the ``standard'' $\Lambda\rightarrow n+\pi$
mechanism. It is interesting to consider the value of the effect in other light hypernuclei. To be specific we
consider ${}^{5}_{\Lambda}$He. Within the same assumptions (absence of the wave function suppression
related to occurrence of a core in the $\Lambda$-core potential)  one arrives to the same ratio
of the ``absorption'' and ``charge-exchange'' amplitudes as (\ref{ratio1}) or, equivalently, (\ref{cross}). 
In this case, however, the cross-sections entering in (\ref{cross}) are those of the 
$\pi^0+{}^{4}{\rm He}\rightarrow \pi^{+} + X$ and $\pi^{-}+{}^{4}{\rm He}\rightarrow  X$ processes.

The low-energy (${\rm T}_{\rm lab}\approx 60$~MeV) value of the ${}^{4}$He absorption cross section equals
\cite{Fowler}:
\begin{equation}
\sigma^{\rm abs}\approx 30\;{\rm mb}
\label{abs4}
\end{equation}

The value of the $\pi$-meson charge exchange cross section on the ${}^{4}$He nucleus 
at ${\rm T}_{\rm lab}= 120$~MeV is \cite{Baum}:
\begin{equation}
\sigma^{\rm cex}\approx 14\;{\rm mb}
\label{cex4}
\end{equation}
The $\pi$-meson energy in the experiment \cite{Baum} corresponds to the resonance region. It is reasonable to expect
that in the low-energy region we are interested in, i.e. beyond the resonance, the matrix element
of the charge-exchange reaction will be smaller. Additional suppression comes from the
smallness of the phase space volume, since the minimal final-state configuration contains
3 particles (e.g. $\pi^+ + n +T$).

Substituting the (\ref{cex4}) and (\ref{abs4}) values of the cross sections in the ratio (\ref{cross}) we get
the estimate of the relative contribution of two mechanisms (``absorption'' and ``charge-exchange'') in the
$\pi^{+}$-emission from the ${}^{5}_{\Lambda}$He nucleus:
\begin{equation}
\frac{M^{\pi\pi}}{M^{\pi}}\agt 1
\end{equation}
It shows that in the ${}^{5}_{\Lambda}$He decay two mechanisms give 
the comparable contribution as it is the case of the ${}^{4}_{\Lambda}$He nucleus. 

The main difference between two nuclei (${}^{5}_{\Lambda}$He and ${}^{4}_{\Lambda}$He) is that in the former case 
the  minimal-configuration 
``charge-exchange'' decay channel contains 3 particles in contrast to two particles
(${}^{4}_{\Lambda}{\rm He}\rightarrow \pi^{+} + {}^3{\rm H}$) in the  latter case. 
Since the energy yield is small, the probability of the ${}^{5}_{\Lambda}{\rm He}\rightarrow \pi^{+} + X$ decay
will be essentially smaller than that of the  ${}^{4}_{\Lambda}{\rm He}\rightarrow \pi^{+} + X$ decay.
Hence the contribution of both mechanisms will be kinematically suppressed and both  $\pi^+$ decay
channel will be hardly observable in this case. 

The situation changes in the case of heavier hypernuclei. The relative contribution of two mechanisms is
controlled again by the ratio of the ``charge-exchange'' and ``absorption'' cross section on the nuclear core.
For Li$\rightarrow$B nuclei the charge exchange channel makes $7\rightarrow 5\%$ \cite{Ing} of the
the total cross section  while the contribution of the absorption cross section is $22\rightarrow 35\%$ \cite{Ing}.
One can conclude safely that in heavy hypernuclei the ``absorption'' mechanism dominates over the
``charge-exchange'' one by a factor of $3\div 7$. For the majority of nuclear cores in this mass interval the 
minimal-configuration
charge-exchange channel $\pi^{0}+{\rm (A,Z)}\rightarrow \pi^{+} +{\rm (A, Z-1)}$ exists. It will
give a contribution to the decay probability comparable to that of the 
$\pi^{0}+{}^3{\rm He}\rightarrow \pi^{+} + {}^3{\rm H}$ channel in the ${}^{4}_{\Lambda}$He decay.
Then the ``absorption'' mechanism will yield even larger contribution enhancing the $\pi^{+}$-decay
modes. 
\section{Conclusions}
We considered a mechanism of $\pi^{+}$ hypernuclear decay generated by the weak
$\Lambda\rightarrow n + \pi^{-} + \pi^{+}$ vertex  with the subsequent absorption of the $\pi^{-}$-meson
on the nuclear core Its relative contribution compared to that of the standard mechanism 
($\Lambda\rightarrow n+\pi^{0}$ decay followed by the $\pi^{0}\rightarrow\pi^{+}$ charge-exchange)
is proportional to the ratio of the cross section of  the $\pi^-$ absorption by the core and of the charge exchange 
scattering.
This ratio in the case of the ${}^{4}_{\Lambda}{\rm He}\rightarrow \pi^{+}$ decay is close to unity. It implies
that the account of the $\Lambda\rightarrow n + \pi^{-} + \pi^{+}$ mechanism can yield a plausible solution of the
challenging problem of the $\pi^{+}$ meson emission probability. The role of the ``absorption'' mechanism is
expected to be even more pronounced in heavy hypernuclei where it will be a dominant mechanism of the
$\pi^+$ emission. This conclusion makes remeasurement of the $\pi^{+}$ emission probability highly
desirable.

The main ambiguity of the predicted effect is related  to the off-shell kinematics of the absorbed $\pi^-$-meson.
Generically, the cross section of the process may be expanded in the number of nucleons involved:
\begin{equation}
\sigma^{\rm abs} = \sigma^{\rm abs}_{\rm N}+ \sigma^{\rm abs}_{\rm 2N} + ...
\label{nn}
\end{equation}
For the on-shell pion the first term $\sigma^{\rm abs}_{\rm N}$ vanishes and the total cross section
is dominated by the two-nucleon term. The situation changes when the absorbed $\pi$-meson is
off-shell. Then the first term (\ref{nn}) has a nonvanishing value, while the $\sigma^{\rm abs}_{\rm 2N}$ term
becomes suppressed due to  formfactors. Accurate solution of the problem requires rigorous
analysis of a multinucleon system. One might expect. however, that for the $\pi$-meson, which is
slightly off-shell, the situation does not differ drastically from that for the on-shell $\pi$-meson and
all the conclusions inferred on the basis of the comparison of the charge-exchange and absorption
cross sections for real $\pi$-mesons hold true.

The suggested mechanism has also some qualitative features which are in line with experimental
observations. First, the $\pi$-mesons are emitted in the $\Lambda\rightarrow n + \pi^{-} + \pi^{+}$ vertex in the
S-wave. Second, the $\pi^{-}$-meson is off-shell and the probability that it will reach the nuclear core
depends upon its virtuality. The mass squared $q^2$ of the  $\pi^{-}$-meson is related to the effective mass
of the $\pi^{+} + n$ pair $\mu$: $q^2 = m^2_{\Lambda} - 2m_{\Lambda}\,\mu$. It shows that the smaller is
the effective mass $\mu$ the smaller is the $\pi^{-}$-meson virtuality. The $\mu$ diminishes (and virtuality
gets smaller)  for small energy of the emitted $\pi^{+}$-meson. Thus the ``absorption'' mechanism populates the
low-energy domain of the $\pi^{+}$-meson spectrum. These features correspond to experimental observations
\cite{Bohm}. Third, absorption of the $\pi$-meson involves (for the on-shell pion) at least two nucleons (and it is
natural to expect that this feature survives for the pion not far from the mass shell). Again multinucleon final states
are also in line with experimental results \cite{Mayeur}.\\[0.15cm]
\begin{center}
{\bf Acknowledgements}
\end{center}
The author is indebted to A.Gal, M.Oka and A.Ramos for helpful discussions.

\newpage
\unitlength=1mm
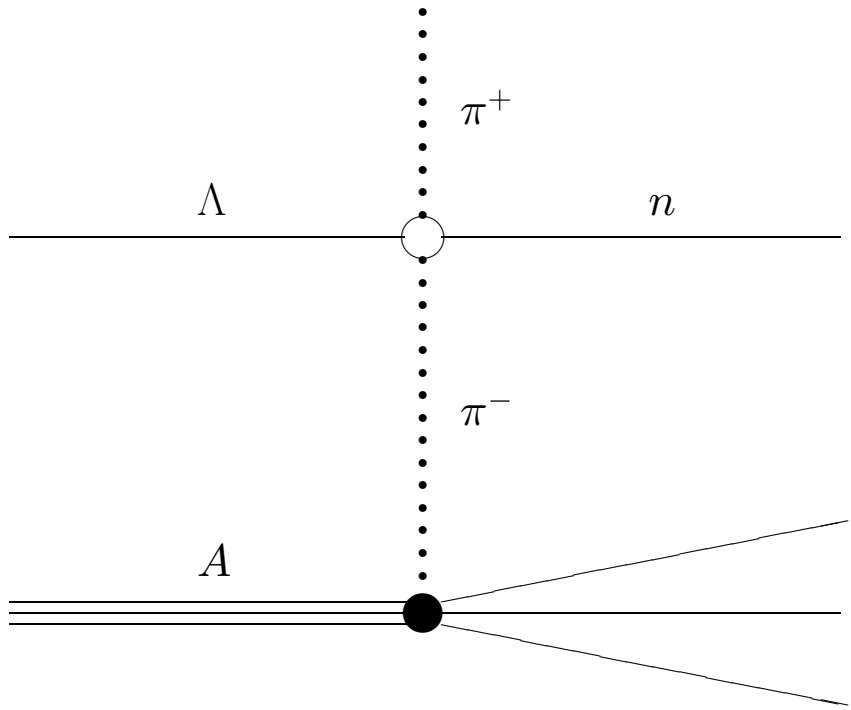
\begin{figure}
\begin{picture}(100,120)
\put(60,60){\circle{6}}
\put(5,60){\line(1,0){52.5}}
\put(62.5,60){\line(1,0){53}}
\put(60,10){\circle*{20}}
\put(5,8.5){\line(1,0){53}}
\put(5,10){\line(1,0){53}}
\put(5,11.5){\line(1,0){53}}
\put(62.5,8.5){\line(5,-1){54}}
\put(62.5,10){\line(1,0){53}}
\put(62.5,11.5){\line(5,1){54}}
\multiput(60,63)(0,3){10}{\circle*{1}}
\multiput(60,57)(0,-3){15}{\circle*{1}}
\put(30,63){{\Large $\Lambda$}}
\put(90,63){{\Large$n$}}
\put(65,75){{\Large$\pi^+$}}
\put(65,35){{\Large$\pi^-$}}
\put(30,15){{\Large $A$}}
\end{picture}\\[0.4cm]
\caption{ Diagram of the $\Lambda\rightarrow n+\pi^{+} +\pi^-$ weak decay (empty circle) 
followed by the $\pi^-$-meson absorption by the hypernuclear core $A$ (filled circle). Solid 
and dotted line denote baryons and $\pi$-mesons respectively}
\end{figure}
\newpage
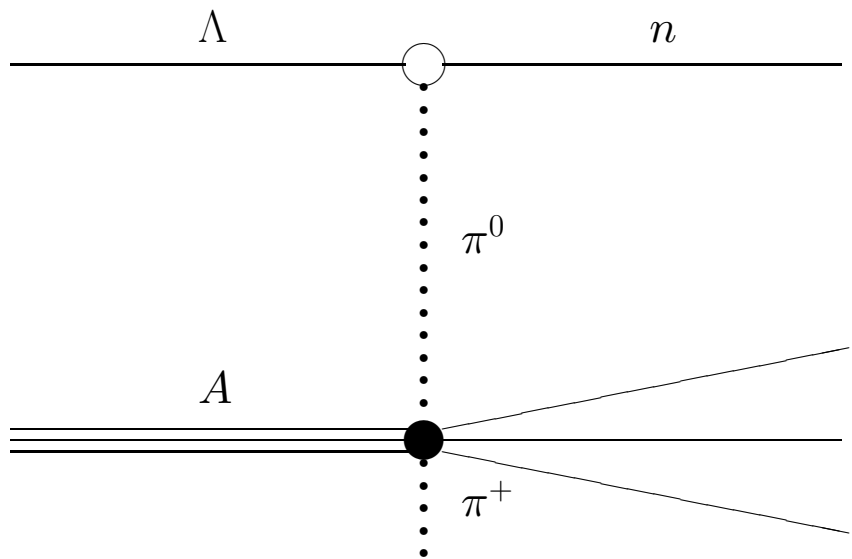
\begin{figure}
\unitlength=1mm
\begin{picture}(100,120)
\put(60,60){\circle{6}}
\put(5,60){\line(1,0){52.5}}
\put(62.5,60){\line(1,0){53}}
\put(60,10){\circle*{6}} 
\put(5,8.5){\line(1,0){53}}
\put(5,10){\line(1,0){53}}
\put(5,11.5){\line(1,0){53}}
\put(62.5,8.5){\line(5,-1){54}}
\put(62.5,10){\line(1,0){53}}
\put(62.5,11.5){\line(5,1){54}}
\multiput(60,57)(0,-3){15}{\circle*{1}}
\multiput(60,7)(0,-3){5}{\circle*{1}}
\put(30,63){{\Large $\Lambda$}}
\put(90,63){{\Large$n$}}
\put(65,0){{\Large$\pi^+$}}
\put(65,35){{\Large$\pi^0$}}
\put(30,15){{\Large $A$}}
\end{picture}\\[0.4cm]
\caption{Diagram of the $\Lambda\rightarrow n+\pi^{0}$ weak decay (empty circle)
followed by the $\pi^0\rightarrow\pi^+$ charge exchange on the hypernuclear core (filled circle)}
\end{figure}

\begin{thebibliography}{99}
\bibitem{Bohm} G.Bohm {\it et al}., Nucl. Phys. {\bf B9}, 1 (1969).
\bibitem{Keyes}G.Keyes {\it et al}, Nuovo Cimento {\bf 31A}, 
401 (1976).
\bibitem{Dalitz} R.H.Dalitz and F. von Hippel, Nuovo Cimento {\bf 34}, 799 (1964).
\bibitem{Hippel} F. von Hippel, Phys. Rev.{\bf 136}, B 455 (1964).
\bibitem{Gal} A.Ciepl\'y and A.Gal, Phys.Rev. C {\bf 55}, 2715 (1997).
\bibitem{Gibson} B.F.Gibson and R.G.E.Timmermans, nucl-th/9711054.
\bibitem{Oka} M.Oka, nucl-th/9711049.
\bibitem{Donoghue} J.F.Donoghue, E.Golowich, and B.R.Holstein, Dynamics of the Standard
Model, {\it Cambridge University Press}, 1992
\bibitem{sp} M.A.Khandaker {\it et al}., Phys. Rev. C {\bf 44}, 24 (1991).
\bibitem{abs} P.Weber {\it et al}., Nucl. Phys. {\bf A534}, 541 (1991).
\bibitem{Fowler}E.C.Fowler {\it et al}.,
Phys.Rev. {\bf 91}, 135 (1953).
\bibitem{Baum}M.Baumgartner {\it et al}., Nucl. Phys. {\bf A399}, 451 (1983).
\bibitem{Ing}C.H.Q.Ingram, Nucl. Phys. {\bf A374}, 319c (1982).
\bibitem{Mayeur} C.Mayuer {\it et al}., Nuovo Cim. {\bf 44}, 698 (1966)
\end{thebibliography}
\end{document}